\documentclass[preprint,12pt]{elsarticle}

\usepackage{amssymb}

\usepackage{amsfonts}
\usepackage{latexsym}
\usepackage{amsmath,epsfig,algorithmic,color,balance}
\usepackage[caption=false]{subfig} 
\usepackage{multirow}
\usepackage{booktabs}

\providecommand{\abs}[1]{\lvert#1\rvert}
\providecommand{\norm}[1]{\lVert#1\rVert}
\newtheorem{theorem}{Theorem}
\newcommand{\argmin}{\operatornamewithlimits{argmin}}

\journal{Applied and Computational Harmonic Analysis}

\begin{document}

\begin{frontmatter}

\title{Optimized projections for compressed sensing via rank-constrained nearest correlation matrix}

\author{Nicolae Cleju}

\address{``Gheorghe Asachi'' Technical University of Iasi, Romania}

\begin{abstract}
Optimizing the acquisition matrix is useful for compressed sensing of signals that are sparse in overcomplete dictionaries, because the acquisition matrix can be adapted to the particular correlations of the dictionary atoms. In this paper a novel formulation of the optimization problem is proposed, in the form of a rank-constrained nearest correlation matrix problem. Furthermore, improvements for three existing optimization algorithms are introduced, which are shown to be particular instances of the proposed formulation. Simulation results show notable improvements and superior robustness in sparse signal recovery.
\end{abstract}


\end{frontmatter}

\section{Introduction}
\label{sec_intro}

Compressed Sensing (CS) \cite{Donoho2006CS} studies the possibility of acquiring a signal $x$ that is a priori known to be sparse in some dictionary $D$ with fewer linear measurements than required by the traditional sampling theorem. In many cases the dictionary $D$ is an orthogonal basis, but we consider here the general case of an overcomplete dictionary.

Consider a signal $x \in \mathbb{R}^n$ that is sparse in some dictionary $D \in \mathbb{R}^{n \times N}$, i.e $x$ has at least one decomposition $\gamma$ that has few non-zero coefficients. A number of $m < n$ linear measurements are taken as inner products of $x$ with a set of $m$ projection vectors, arranged as the rows of an acquisition matrix $P \in \mathbb{R}^{m \times n}$
\begin{equation}
\displaystyle
y = P x = \underbrace{P D}_{D_e} \gamma.
\label{eq_sense}
\end{equation}
The equation system (\ref{eq_sense}) is undetermined. Under certain conditions on $P$ and $D$ \cite{CSStableSigRec}, a sufficiently sparse decomposition vector $\gamma$ is shown to be the unique solution to the optimization problem 
\begin{equation}
\displaystyle
\hat{\gamma} = \arg\min_\gamma \norm{\gamma}_{\ell_0} \text{ subject to } y = PD \gamma,
\label{eq_rec_L0}
\end{equation}
where $\|\gamma\|_{\ell_0}$ is the number of non-zero elements of the vector $\gamma$ (the $\ell_0$ ``norm''). Solving (\ref{eq_rec_L0}) means finding the sparsest decomposition of $y$ in the effective dictionary $D_e := P D$, which is the computational expensive stage of the process, with a large number of algorithms developed for this purpose. After obtaining the approximate decomposition vector $\hat{\gamma}$, the reconstructed signal $\hat{x}$ is obtained as 
\begin{equation}
\displaystyle
\hat{x} = D \hat{\gamma}.
\label{eq_recx}
\end{equation}
The strict condition $y = PD \gamma$ in \eqref{eq_rec_L0} is often unrealistic, and therefore a practical version of \eqref{eq_rec_L0} is 
\begin{equation}
\displaystyle
\hat{\gamma} = \arg\min_\gamma \norm{\gamma}_{\ell_0} \text{ subject to } \norm{y - PD \gamma} \leq \epsilon
\label{eq_rec_L0_approx}
\end{equation}
where $\epsilon$ takes into account possible noisy measurements and approximately sparse signals.

Unfortunately, finding the exact solution of the $\ell_0$ minimization problem \eqref{eq_rec_L0} is combinatorial and NP-hard. One of the ways to circumvent this is replacing the $\ell_0$ norm with $\ell_1$, leading to a tractable convex optimization problem 
\begin{equation}
\displaystyle
\hat{\gamma} = \arg\min_\gamma \norm{\gamma}_{\ell_1} \text{ subject to } y = P D \gamma,
\label{eq_BP}
\end{equation}
which requires however more strict conditions on $P$ and $D$ to guarantee the uniqueness of the solution. This is known as Basis Pursuit (BP) \cite{DecodingByLPCandes2005}. The $\ell_1$ problem can be converted to a linear program, which is well known in literature and has many efficient solving algorithms available. A second option is to settle with a possibly sub-optimal solution of (\ref{eq_rec_L0}), using a pursuit or thresholding   algorithm \cite{Pati1993, Blumensath2009} to estimate a solution to \eqref{eq_rec_L0}. In both cases, robustness to noise can be enforced by replacing the strict condition $y = P D \gamma$ with a robust $\norm{y - P D \gamma}_2 \le \epsilon$.

The choice of the acquisition matrix $P$ is governed by the principle of incoherence with $D$: a ``good'' acquisition matrix has its rows (i.e. the projection vectors) incoherent with the columns of $D$. Coherence measures the largest correlation between two sets of vectors, and thus incoherence requires a low maximal correlation. Random projections vectors were shown to be a good choice with orthogonal bases \cite{Candes2006RandProj}, since random vectors are incoherent with any fixed basis with high probability. In the overcomplete case, a better acquisition matrix can often be found if one takes into account the correlations between dictionary atoms, since it is not uncommon that dictionaries exhibit significant atom correlation. This is especially true with dictionaries that are learned, i.e. optimized for a particular set of signals. As such, a number of algorithms have been developed for finding optimized projections for signals that are sparse in overcomplete dictionaries \cite{EladOptimProj, Xu, Duarte}.

This paper proposes modifications for improving three existing algorithms for finding optimized projections. Further, we show that our improvements can be unified in a single formulation based on solving a \emph{rank-constrained nearest correlation matrix} problem \cite{GaoMajPen}. The rest of this paper is organized as follows. In Section \ref{sec_proj} we review the main condition for perfect recoverability that underlies most of the considered algorithms. Section \ref{sec_algos} presents three state-of-the-art algorithms for finding optimized projections. Improvements for all of them are proposed in Section \ref{sec_improving}, and we present the proposed unified formulation in Section \ref{sec_proposed}. Simulation results are presented in Section \ref{sec_res}. Finally, conclusions are drawn in Section \ref{sec_concl}.

Throughout this paper we use the following notations. The acquired signal is an $n$-dimensional vector $x$, the dictionary is $D$ of size $n \times N, n < N$. A decomposition of $x$ in $D$ is typically denoted as $\gamma$, i.e. $x = D \gamma$. The acquisition matrix is $P$ of size $m \times n, m < n$. The product $D_e := P D$ is the effective dictionary, of size $m \times N$. The Gram matrix of $D$ is denoted $G := D^T D$, while the Gram matrix of the effective dictionary $D_e$ is denoted $G_e$ and referred to as \emph{effective Gram matrix}.

\section{Acquisition matrices and mutual coherence}
\label{sec_proj}

A widely used approach to ensure the uniqueness of the solution $\hat{\gamma}$ in (\ref{eq_rec_L0}) or (\ref{eq_BP}) uses the mutual coherence of the effective dictionary $D_e := P D$. The mutual coherence of a dictionary is defined as the maximum absolute value of the inner products of any two of its normalized columns \cite{DonohoElad2003}. Thus, the mutual coherence of $D_e$ is the maximum absolute off-diagonal value of the Gram matrix $G_e := D_e^T \cdot D_e$, after normalizing the columns of $D_e$. The mutual coherence provides a lower bound for the perfect recovery of sparse signals, as shown in Theorem \ref{th_MutCoh} \cite{DonohoElad2003,Gribonval2003,TroppGreed}:
\begin{theorem}
\label{th_MutCoh}
Consider an overcomplete dictionary $D$ with mutual coherence $\mu(D)$ and a signal $x$ such that $x = D \gamma$. If condition (\ref{eq_mutCond}) is true:
\begin{equation}
\label{eq_mutCond}
|| \gamma ||_0 < \frac{1}{2} \left( 1 + \frac{1}{\mu(D)} \right)
\end{equation}
then the following hold:
\begin{enumerate}
\item $\gamma$ is the sparsest decomposition of $x$ in $D$, i.e. it is the solution of the optimization problem \[\argmin_\gamma \| \gamma \|_0 \text{ subject to } x = D \gamma\]
\item $\gamma$ is recoverable using $\ell_1$ minimization \cite{DecodingByLPCandes2005}, i.e. it is also the solution of \[\argmin_\gamma \| \gamma \|_1 \text{ subject to } x = D \gamma\]
\item $\gamma$ is recoverable using Orthogonal Matching Pursuit \cite{Pati1993}.
\end{enumerate}
\end{theorem}

Theorem \ref{th_MutCoh} shows that having a smaller mutual coherence of the dictionary is desirable, as it increases the set of recoverable signals. As such, an optimal acquisition matrix $P$ is one that minimizes the mutual coherence of the effective dictionary $D_e := P D$, i.e. the largest off-diagonal element of the effective Gram matrix $G_e := D_e^T D_e$. The optimization problem can be stated in terms of minimizing the mutual coherence, or, in general, the largest off-diagonal elements of $G_e$.

\section{Existing optimization algorithms}
\label{sec_algos}

\subsection{The algorithm of Elad}
\label{subsec_Elad}

The algorithm of Elad \cite{EladOptimProj} aims to reduce the $t$-averaged mutual coherence $\mu_t$ of the effective dictionary $D_e$, defined as the average of the $t$ largest off-diagonal values of the Gram matrix: 
\begin{equation} \displaystyle
\label{eq_mutDe}
\mu_t(D_e) = \frac{   \sum_{1 \le i,j \le k, i \neq j}\left( |g_{ij}| > t \right) \cdot |g_{ij}|  }{   \sum_{1 \le i,j \le k, i \neq j}\left( |g_{ij}| > t \right)   }
\end{equation}
The parameter $t$ is either a fixed threshold or a percentage indicating the top fraction of the matrix elements that are to be considered. The reason for minimizing the $t$-averaged value instead of the single largest off-diagonal value is that the latter is a pessimistic bound: even under more relaxed conditions than (\ref{eq_mutCond}), in practice almost all signals can still be adequately recovered, at the expense of a small fraction of unrecoverable signals. For this reason, it is argued that the $t$-averaged mutual coherence is a better measure for the \emph{average} behavior of the effective dictionary.

The algorithm iteratively shrinks the off-diagonal elements of the effective Gram matrix $G_e$, while keeping the rank of the matrix equal to $m$. At every iteration $k$, the largest off-diagonal values of the current Gram matrix $G_e^{(k)}$ are reduced using a shrinking function $f_t(u)$. This is followed by a low rank approximation to enforce the required rank $m$. The resulting effective dictionary $D_k$ is presumably better than the initial one, as it has smaller mutual coherence. The acquisition matrix at step $k$, denoted as $P_k$, is then found as $P = D_k D^\dagger$, (where $^\dagger$ denotes the Moore-Penrose pseudoinverse).

The shrinking function is empirically chosen as in \eqref{eq_shrfun}, for some parameter $\alpha < 1$.

\begin{equation} \displaystyle
\label{eq_shrfun}
f_t(g_{ij}) = 
	\begin{cases}
		g_{ij}                            &  |g_{ij}| \leq \alpha t \\
		\alpha t \cdot sgn(g_{ij}) &  \alpha t \leq |g_{ij}| \le t \\
		\alpha g_{ij}                   & t \le |g_{ij}|
	\end{cases}
\end{equation}

The complete algorithm is summarized in Fig.\ref{algo_Elad}.

\begin{figure}
\begin{algorithmic}[1]
  \REPEAT
		\STATE Compute the effective dictionary $D_e = P_k \cdot D$, normalize its columns, and compute its Gram matrix $G_e^{(k)} = D_e^T \cdot D_e$
		\STATE Apply shrinking function to the off-diagonal elements of $G_e^{(k)}$,  $\hat{G_e}^{(k)} = f \left( G_e^{(k)} \right)$
		\STATE Find the best rank $m$ approximation of $\hat{G_e}^{(k)}$ using singular value decomposition
		\STATE Extract square root $D_k$, where $\hat{G_e}^{(k)} = D_k^T \cdot D_k$
		\STATE Choose $P_k = D_k D^{\dagger}$ , i.e. minimizing $||D_k - P_k \cdot D||_F$
	\UNTIL{Until stop criterion}
\end{algorithmic}
\caption{The Elad algorithm}
\label{algo_Elad}
\end{figure}

\subsection{The algorithm of Xu \emph{et al}}
\label{subsec_Xu}

The algorithm of Xu \emph{et al} \cite{Xu} aims to make the effective dictionary $D_e$ as close as possible to an equiangular tight frame (ETF), because an ETF has minimal mutual coherence among all matrices of the same dimension. Thus, it aims to solve the optimization problem
\begin{equation}
\hat{G_e} = \min_{G \in A_m^n } \norm{G_e - G}
\end{equation}
where $A_m^n$ is the set of the Gram matrices of all $m \times n$ ETFs. Since this set is not convex, they replace it with the convex set $\Lambda^n$:
\begin{equation}
\begin{split}
\Lambda^n = \{ G \in \mathbb{R}^{n \times n}: \; & G = G^T, \text{diag}(G)=1,\\
 & \max_{i \neq j}\|g_{ij}\| < \mu_G \}
\end{split}
\end{equation}
where $\mu_G = \sqrt{\frac{n-m}{m(n-1)}}$ is a lower bound for the coherence of an $m \times n$ ETF. The term $\text{diag}(G)$ refers to the main diagonal of $G$, and thus the condition $\text{diag}(G)=1$ requires all elements on the main diagonal to be equal to 1. 
The algorithm, based on alternating projections, is presented in Fig.\ref{algo_Xu}.

Note that the Xu algorithm is very similar to the Elad algorithm, up to a different choice of the shrinkage function for the largest elements, and with a reduced step taken towards the prospective solution at every iteration. The value of the step size $\alpha$, however, is not given, nor is any suggestion on how to choose an adequate value. As such, in our simulations we test a range of values of $0.1, 0.2 ... 1.0$ and aggregate the best results.

\begin{figure}
\begin{algorithmic}[1]
  \REPEAT
		\STATE Compute the effective dictionary $D_e = P \cdot D$, normalize its columns and compute the Gram matrix $G_e^{(k)} = D_e^T \cdot D_e$
		\STATE Project $G_e$ on $\Lambda^k$ by enforcing:
			\begin{equation} \displaystyle
			\label{eq_shrXu}
			g_{ij} = 
			\begin{cases}
				1                    & i = j \\
				g_{ij}               & \abs{g_{ij}} < \mu_G \\
				sgn(g_{ij}) \cdot \mu_G   & \abs{g_{ij}} \geq \mu_G
			\end{cases}
			\end{equation}
		\STATE New solution is between the projection $G_P$ and the previous solution:
			\begin{equation}
			G_k = \alpha G_P + (1-\alpha) G_{k-1}\;\;, \;\; 0 < \alpha < 1
			\end{equation}
		\STATE Update the acquisition matrix P using QR factorization with eigenvalue decomposition 
	\UNTIL{Until stop criterion}
\end{algorithmic}
\caption{Xu algorithm}
\label{algo_Xu}
\end{figure}

\subsection{The algorithm of Duarte-Carvajalino and Sapiro}
\label{subsec_Duarte}

Duarte-Carvajalino \& Sapiro introduce in \cite{Duarte} a different algorithm for finding optimized projections for a given dictionary, as well as a method for joint dictionary and acquisition matrix optimization. Since we consider fixed dictionaries, we will focus only on the former, which we refer to as the \emph{Duarte algorithm} for brevity. The authors seek the acquisition matrix $P^\star$ that minimizes:
\begin{equation} \displaystyle
\label{eq_Duarte2}
P^\star = \argmin_P ||D D^T - D D^T P^T P D D^T||_F
\end{equation}
The problem \eqref{eq_Duarte2} has a closed-form solution in the form
\begin{equation} \displaystyle
\label{eq_DuarteP}
P^\star = \Lambda_{1:m}^{-1/2} \cdot U_{1:m}^T
\end{equation}
where $\Lambda$ and $U$ come from the eigenvalue decomposition of $D D^T = U \Lambda U^T$ and the notation $_{1:m}$ indicates a restriction to the first $m$ eigenvectors and eigenvalues. In other words, considering a singular value decomposition (SVD) of $D = U S V^T$, the optimal acquisition matrix is given by the top $m$ principal components of $D$ scaled with the inverse of the corresponding singular values, $P^\star = S_{1:m}^{-1} U_{1:m}^T$. As a consequence, the resulting effective dictionary is a tight frame defined by the restriction of right singular matrix $V^T$ to the top $m$ rows:
\begin{equation} \displaystyle
\label{eq_DuarteDe}
D_e = P^\star \cdot D = S_{1:m}^{-1} U_{1:m}^T \cdot U S V^T = V_{1:m}^T
\end{equation}

\section{Improving existing optimization algorithms}
\label{sec_improving}

\subsection{Improving the Elad and Xu algorithms}
\label{subsec_improvingEladXu}

\subsubsection{Reformulating as constrained optimization}

As a first step towards improving the existing algorithms, we reformulate them as constrained optimization problems. 

We first note that the Elad algorithm can be thought of as a way of robustly solving the following optimization problem:
\begin{equation}
\label{eq_EladAlgoRef}
\begin{split}
\textbf{minimize } & \norm{G_e - I_N}_\infty \\
\textrm{subject to: } & G_e \succeq 0 \\ 
& \text{rank}(G_e) = m \\ 
& \text{diag}(G_e) = 1
\end{split}
\end{equation}
Indeed, at every iteration the algorithm shrinks the largest off-diagonal elements of $G_e$, which effectively is a robust way of reducing $\norm{G_e - I_N}_\infty$,  followed by enforcing the rank, positive semidefiniteness  and unit-diagonal constraint. Thus, the Elad algorithm can be considered an iterative method for constrained $\ell_\infty$ minimization.

A similar reasoning holds for the Xu algorithm. It is well known that the maximum correlation of two atoms is minimal for an ETF among all other matrices of same size \cite{Welch, ETF}. It follows that projecting on the set of ETFs is similar to minimizing the largest absolute off-diagonal value of $G_e$, i.e. $\norm{G_e - I_N}_\infty$. As such, the Xu algorithm can also be thought of as a way of solving the same constrained optimization problem \eqref{eq_EladAlgoRef} as the Elad algorithm. As presented in section \ref{subsec_Xu}, the two algorithms follow the same approach, iteratively shrinking the large off-diagonal elements of $G_e$ and enforcing the constraints, with minor differences in the shrinkage functions.

\subsubsection{Analysis and proposed improvement}

We first analyze the Elad and Xu algorithms in two corner cases, where an intuitive analysis suggests that the optimization problem is too strict. We then propose a relaxed version that not only handles the two particular cases, but provides better results overall, as shown in Section \ref{sec_res}.

In the first scenario, let us consider that the initial acquisition matrix is simply the full-size identity matrix $P = I_n$. This is obviously the ideal case, as the acquired vector $y$ is exactly the desired signal $x$. There is nothing to optimize in this case, the acquisition matrix $P$ is perfect. However, both Elad and Xu algorithms fail to notice this. The effective dictionary $D_e$, being identical to $D$, still has the same coherence as the atoms of $D$. The two algorithms proceed to shrink the large off-diagonal values of the Gram matrix $G_e$ as usual, failing to recognize that the acquisition matrix is already optimal.

In a second scenario, let us consider that the dictionary $D$ contains two identical atoms $d_i$ and $d_j$ (even though this is an extreme scenario, we use it only as a toy example to illustrate the shortcomings when dealing with linear dependency between atoms). It follows that the effective dictionary $D_e := P D$ will also have two identical columns $d_{e_i}$ and $d_{e_j}$, for any acquisition matrix $P$. This means maximal mutual coherence. The effective Gram matrix $G_e$ will therefore have a pair of $1$'s outside the main diagonal, which the two algorithms will strive to reduce. However, this is unnecessary:  even if the sparse decomposition is not unique (atom $i$ and atom $j$ can be swapped because of the ambiguity between them), the reconstructed signal $x$ is actually the same irrespective of which atom is used, since the two corresponding atoms in $D$ are identical as well. We may think of this as \emph{correlations in the effective dictionary $D_e$ that are inherited from the original dictionary $D$ are not bad}. In this case the optimization algorithm should not worry about the off-diagonal $1$'s of the Gram matrix, since any ambiguity in deciding which of the two atoms to use is irrelevant when it comes to reconstructing the signal from the atoms of $D$. 

A solution to both problems is to replace the minimization of $\norm{G_e - I_N}$ with $\norm{G_e - G}$. This solves the above shortcomings: in the first case, $G_e = G$ and the algorithm recognizes it is already optimal, whereas having two identical columns in both $D$ and $D_e$ means a a pair of off-diagonal zeroes in $G_e - G$, indicating there is nothing to optimize for the two atoms.

We propose therefore to pose the optimization problem as:
\begin{equation}
\label{eq_EladAlgoModif}
\begin{split}
\textbf{minimize } & \norm{G_e - G}_\infty \\
\textrm{subject to: } & G_e \succeq 0 \\ 
& \text{rank}(G_e) = m \\ 
& \text{diag}(G_e) = 1
\end{split}
\end{equation}
Consequently, we propose the modification of Elad and Xu algorithms by making them reduce the largest off-diagonal values of the difference $G_e - G$, instead of $G_e$. We refer to these algorithms as RCNCM-Elad and RCNCM-Xu, and their complete description is given in Section \ref{sec_proposed}. The acronym RCNCM stands for rank-constrained nearest correlation matrix, for reasons that are explained in Section \ref{sec_proposed}.

Note that in the orthonormal case the proposed modification reduces to the original problem, since $D$ being an orthonormal basis implies $G = I_N$ and \eqref{eq_EladAlgoModif} becomes identical to \eqref{eq_EladAlgoRef}.

\subsection{Improving the Duarte algorithm}
\label{subsec_improvingDuarte}

\subsubsection{Reformulating as constrained optimization}
The Duarte algorithm is originally formulated as an optimization problem seeking the minimization of \eqref{eq_Duarte2}. For consistency with the other two algorithms, we prefer to reformulate the optimization problem as:
\begin{equation}
\label{eq_DuarteOptimProbRef}
\begin{split}
\textbf{minimize } & \norm{G_e - G}_2 \\
\textrm{subject to: } & G_e \succeq 0 \\ 
& \textrm{rank}(G_e) = m \\ 
\end{split}
\end{equation}
The solution to \eqref{eq_DuarteOptimProbRef} is identical with the solution of the original Duarte problem \eqref{eq_Duarte2}, up to normalization of the projection vectors. Indeed, the solution to \eqref{eq_DuarteOptimProbRef} is given by the Eckart--Young theorem by keeping the most significant $m$ eigenvectors and values of $G$. Considering an SVD factorization of $D = U S V^T$, then $G = D^T D = V S^2 V^T$, the optimal $G_e^\star = V_{1:m} S_{1:m}^2 V_{1:m}^T$, implying $D_e^\star = S_{1:m} V_{1:m}^T$, leading to the optimal acquisition matrix being 
\begin{equation}
\label{eq_DuarteRefSol}
P^\star = U_{1:m}^T
\end{equation}
since $D_e^\star = P^\star D.$ This is essentially the same as the Duarte solution \eqref{eq_DuarteP}, the only difference being that the projection vectors in \eqref{eq_DuarteRefSol} are normalized. However, the scaling of the acquisition vectors plays little role in practice. Thus, one can think of the Duarte algorithm as essentially a way of solving \eqref{eq_DuarteOptimProbRef} followed by scaling the projection vectors.

\subsubsection{Analysis and proposed improvement}
\label{sec_Du_analysis}
Comparing with \eqref{eq_EladAlgoRef} reveals an essential condition missing from \eqref{eq_DuarteOptimProbRef} as well as from the original Duarte formulation: there is no guarantee that the atoms of the effective dictionary $D_e$ are normalized, i.e. $\textrm{diag}(G_e) = 1$. The resulting effective dictionary \eqref{eq_DuarteDe} is composed of the top $m$ rows of the unitary matrix $V^T$, and therefore its atoms have norm smaller than 1. However, norms smaller than 1 artificially reduce the value of the inner products, and therefore minimizing atoms' inner products without ensuring that they are normalized can result in atoms being more coherent than desired (remember that the definition of the mutual coherence requires the atoms of a dictionary to be normalized). In an extreme case some of the atoms in the effective dictionary may be all-zero, meaning that the corresponding atom of $D$ will never be reconstructed, irremediably affecting signal recovery. Note that the Elad and Xu algorithms avoided the problem by explicitly performing a normalization of the effective dictionary at every iteration.

Lack of an atom normalization constraint means that the Duarte optimization problem is less robust in some scenarios, when some atoms of the resulting effective dictionary (top part of the right singular matrix of the dictionary) have very small norms. We provide below a few simple examples:

\begin{description}
\item[Orthonormal basis, unfavorable decomposition.] Consider $D$ any orthonormal basis. An orthonormal basis does not have an unique SVD factorization, but one choice might simply be $D = D I_n I_n$. In this case the optimal Duarte acquisition matrix is $P^\star = D_{1:m}^T$ and the resulting effective dictionary \eqref{eq_DuarteDe} is simply the restriction of the right singular matrix $I_n$ to its top $m$ rows \[D_e = [I_m; 0].\] The last columns of $D_e$ are all-zero, meaning that the corresponding atoms in $D$ cannot be reconstructed from the measurements.

The same thing happens if $D$ is an orthogonal matrix whose atoms are not perfectly normalized (e.g. due to limited precision or noise). The unique SVD is therefore $D = D_n S I_n$, where $D_n$ is the normalized $D$ and $S$ contains the atom norms. Again, $D_e$ consists of the top $m$ rows of $I_n$, meaning the last columns are all-zero and the corresponding atoms are lost.

\item[Concatenation of Dirac and Haar bases.] The dictionary $D$ is obtained as the concatenation of the Dirac basis and the Haar wavelet basis for a full-level decomposition. The norms of the effective dictionary atoms have large variations, a fraction of them being very small.

\item[Non-decimated wavelet dictionary.] $D$ is the non-decimated (i.e. stationary, shift-invariant) dictionary for a 2-level Symmlet4 wavelet decomposition. One third of the atoms in the effective dictionary have very small norms.
\end{description}

The examples above are chosen with no particular purpose other than showing that the Duarte algorithm is distinctively less robust in particular scenarios. Simulation results presented in Section \ref{sec_res} confirm that the signal recovery ratio in these cases is sometimes an order of magnitude below the other algorithms.

We propose therefore a more robust optimization problem that adds the constraint that the effective atoms have unit norm:

\begin{equation}
\label{eq_DuarteModif}
\begin{split}
\textbf{minimize } & \norm{G_e - G}_2 \\
\textrm{subject to: } & G_e \succeq 0 \\ 
& \textrm{rank}(G_e) = m \\ 
\textbf{and } & \text{diag}(G_e) = 1 
\end{split}
\end{equation}
As explained in the next section, this problem is a rank-constrained nearest correlation matrix problem (RCNCM), and can be solved with algorithms developed for robustly estimating correlation matrices \cite{PieterszRankRedCorr,GaoMajPen}. We refer to this problem as RCNCM-Duarte.

\section{Rank-constrained nearest correlation matrix for optimized projections}
\label{sec_proposed}

The considerations in Section \ref{sec_improving} lead us to proposing the following class of optimization problems from choosing the best acquisition matrix:

\begin{equation}
\label{eq_probproposed}
\begin{split}
\textbf{minimize } & \norm{G_e - G}_p \\
\textrm{subject to: } & G_e \succeq 0 \\ 
& \text{rank}(G_e) = m \\ 
& \text{diag}(G_e) = 1 
\end{split}
\end{equation}

This formulation is a natural generalization of all the three proposed algorithms presented above. For $p = 2$, the problem reduces to \eqref{eq_DuarteModif}, i.e. the reformulated Duarte optimization problem with the additional unit-norm constraint. For $p = \infty$, \eqref{eq_probproposed} becomes \eqref{eq_EladAlgoModif} and can be solved with the proposed modifications of the Elad and Xu algorithms introduced in Section \ref{subsec_improvingEladXu}.

The optimization problem \eqref{eq_probproposed} is a rank-constrained nearest correlation matrix problem (RCNCM) \cite{GaoMajPen}. This family of problems has received much attention in the recent years, with applications in finance as well as engineering. A matrix $X$ is called a \emph{correlation matrix} if $X \succeq 0$ (semipositive definite) and $X_{ii} = 1$. In many practical applications, the correlation matrix estimated from noisy, unreliable or possibly incomplete data can turn out to violate the rank and positivity constraints required of a correlation matrix. In these cases, one needs to find a matrix that fulfils the constraints and is close as possible to the input matrix using a distance metric, leading to an optimization problem formulated as in \eqref{eq_probproposed}.

Our interpretation of \eqref{eq_probproposed} is that the effective dictionary should mimic the correlations of the atoms in the original dictionary, by making their Gram matrices as close as possible according to some metric. This approach is intimately related to the overcomplete nature of the dictionary, since the existence of correlations between atoms is automatically implied by overcompleteness.

We consider the $\ell_2$ or the $\ell_\infty$ distances for minimization. Though we give no rigorous justifications in the general case, using the $\ell_\infty$ is justified at least in the orthonormal case: when the dictionary $D$ is an orthonormal matrix, $G = I_N$ and thus  $\norm{G_e - G}_\infty = \norm{G_e - I_N}_\infty$ is the mutual coherence, the minimization of which is guaranteed to improve recovery. No such rigorous guarantee exists for the $\ell_2$ distance, however the improved results reported for the Duarte algorithm \cite{Duarte} support it as a viable option. 

Following these considerations, we name the proposed modified algorithms introduced in the previous section RCNCM-Duarte, RCNCM-Elad and RCNCM-Xu, in order to emphasize their common rank-constrained correlation matrix framework as well as the original algorithm authors.

\subsection{Solving for $p = 2$}
\label{subsec_solvep2}

For $p = 2$, the optimization problem \eqref{eq_probproposed} becomes the modified Duarte problem proposed in \eqref{eq_DuarteModif}. Contrary to the original Duarte algorithm, a simple closed-form solution is not possible due to the additional normalization constraint $\textrm{diag}(G_e) = 1$. This is a rank-constrained nearest correlation matrix problem which has been already studied in the literature, and several approaches have been developed for solving it \cite{PieterszRankRedCorr, GaoMajPen}. 

In this paper we use the majorized penalty approach (MPA) algorithm presented in \cite{GaoMajPen}, based on eigenvalue penalization and majorization, which we summarize here. First, let us note that in absence of the rank constraint, the problem could be formulated as a semidefinite program \cite{GaoLSSDP}. To enforce the additional rank-constraint, the authors of \cite{GaoMajPen} propose to iteratively minimize a function that penalizes the last $(N-m)$ eigenvalues
\begin{equation}
\label{eq_probproposed_1}
\begin{split}
\displaystyle
\textbf{minimize } & f(G_e) = \norm{G_e - G}_2 + c \sum_{m+1}^N \lambda_i \\
\textrm{s.t. } & G_e \succeq 0 \\ 
& \textrm{diag}(G_e) = 1 
\end{split}
\end{equation}
where $c$ is a penalty constant. This is not equivalent to the original problem \eqref{eq_DuarteModif}, but for a large enough value of $c$ the solution of \eqref{eq_probproposed_1} is arbitrarily close to the solution of \eqref{eq_DuarteModif}. Solving \eqref{eq_probproposed_1} is achieved iteratively using the majorization technique \cite{PieterszRankRedCorr}, which consists in solving a sequence of simpler convex optimization problems: given the estimate $G_e^{(k)}$ at iteration $k$, one constructs a simpler convex function $g_k$ that majorizes $f$, $g_k(X) \geq f(X), \; \forall X$, and minimizes $g$ instead, obtaining the new estimate $G_e^{(k+1)}$. The sequence of estimates $G_e^{(k)}$ converges to the solution of \eqref{eq_probproposed_1}. Further details can be found in \cite{GaoMajPen}.

Note that we give only a sketch of the MPA algorithm which omits many details, e.g. the actual majorization function, since for the purposes of this paper we are only interested in it as a way of solving \eqref{eq_DuarteModif}. One can replace the MPA algorithm with any other method for solving \eqref{eq_DuarteModif}.

Once the optimal $G_e$ is found as the solution to \eqref{eq_DuarteModif}, the optimal acquisition matrix $P^\star$ is obtained, as in the other algorithms, by first factorizing $G_e = (D_e^\star)^T D_e^\star$ and then $P^\star = D_e^\star D^\dagger$.

We refer to this algorithm for finding optimized projections as RCNCM-Duarte. It is summarized in Fig.\ref{algo_RCNCM_Du}, including a sketch of the inner MPA algorithm. 

\begin{figure}
\begin{algorithmic}[1]
\STATE Compute Gram matrix $G$ of the dictionary $D$
\STATE Find the solution $G_e^\star$ to the rank-constrained nearest correlation matrix problem \eqref{eq_DuarteModif} using the majorized penalty approach (MPA) algorithm from \cite{GaoMajPen}
\\ \textbf{Start MPA algorithm:}
\REPEAT 
  \STATE Increase (or set initial) penalization constant $c$\\
  \REPEAT[Solve penalized problem \eqref{eq_probproposed_1}]
    \STATE Create majorization function around current solution
    \STATE Minimize the majorized function (convex problem) and update solution
  \UNTIL converged
\UNTIL sum of last $(N-m)$ eigenvalues is below tolerance
\\ \textbf{End MPA algorithm}
\STATE Extract square root $D_e^\star$, where $G_e^\star = (D_e^\star)^T \cdot D_e^\star$
\STATE Choose $P_k = D_k D^{\dagger}$ , i.e. minimizing $||D_k - P_k \cdot D||_F$
\end{algorithmic}
\caption{The RCNCM-Duarte algorithm. The algorithm finds an optimized acquisition matrix from the solution of a rank-constrained nearest correlation matrix problem.}
\label{algo_RCNCM_Du}
\end{figure}

\subsection{Solving for $p = \infty$}
\label{subsec_solvepinf}

For $p = \infty$, the problem \eqref{eq_probproposed} can be solved with the modified versions RCNCM-Elad and RCNCM-Xu introduced in Section \ref{subsec_improvingEladXu}, by iteratively shrinking the top largest off-diagonal elements of the difference matrix $G_e - G$ instead of $G$. The complete description of the two proposed algorithms RCNCM-Elad and RCNCM-Xu is given in Fig.\ref{algo_RCNCM_Elad} and Fig.\ref{algo_RCNCM_Xu}. We point our that although RCNCM-Xu keeps the same projection step in the original Xu algorithm, this has little relation with the original purpose of projecting on the set of ETFs. Instead, we consider it just a different method of reducing the large off-diagonal values of $G_e - G$, using a different shrinking function than in RCNCM-Elad.

\begin{figure}
\begin{algorithmic}[1]
  \REPEAT
		\STATE Compute the effective dictionary $D_e = P_k \cdot D$, normalize its columns, and compute its Gram matrix $G_e^{(k)} = D_e^T \cdot D_e$
		\STATE Compute the difference $\Theta^{(k)} = G_e^{(k)} - G$, where $G = D^T D$
		\STATE Apply shrinking function to the off-diagonal elements of $\Theta^{(k)}$,  $\hat{\Theta}^{(k)} = f \left( \Theta^{(k)} \right)$
		\STATE Compute the estimate matrix by adding back $G$: \\ $\hat{G_e}^{(k)} = \hat{\Theta}^{(k)} + G$
		\STATE Find the best rank $m$ approximation of $\hat{G_e}^{(k)}$ using singular value decomposition
		\STATE Extract square root $D_k$, where $\hat{G_e}^{(k)} = D_k^T \cdot D_k$
		\STATE Choose $P_k = D_k D^{\dagger}$ , i.e. minimizing $||D_k - P_k \cdot D||_F$
	\UNTIL{Until stop criterion}
\end{algorithmic}
\caption{The RCNCM-Elad algorithm}
\label{algo_RCNCM_Elad}
\end{figure}

\begin{figure}
\begin{algorithmic}[1]
  \REPEAT
		\STATE Compute the effective dictionary $D_e = P \cdot D$, normalize its columns and compute the Gram matrix $G_e^{(k)} = D_e^T \cdot D_e$
		\STATE Compute the difference $\Theta^{(k)} = G_e^{(k)} - G$, where $G = D^T D$
		\STATE For $\Theta^{(k)}$ enforce:
			\begin{equation} \displaystyle
			\label{eq_shrXuG}
			\theta_{ij} = 
			\begin{cases}
				1                    & i = j \\
				\theta_{ij}               & \abs{\theta_{ij}} < \mu_G \\
				sgn(\theta_{ij}) \cdot \mu_G   & \abs{\theta_{ij}} \geq \mu_G
			\end{cases}
			\end{equation}
		\STATE Add back $G$: $G_P = \hat{\Theta}^{(k)} + G$
		\STATE New solution is between the projection $G_P$ and the previous solution:
			\begin{equation}
			G_k = \alpha G_P + (1-\alpha) G_{k-1}\;\;, \;\; 0 < \alpha < 1
			\end{equation}
		\STATE Update the acquisition matrix P using QR factorization with eigenvalue decomposition 
	\UNTIL{Until stop criterion}
\end{algorithmic}
\caption{The RCNCM-Xu algorithm}
\label{algo_RCNCM_Xu}
\end{figure}

\section{Simulation results}
\label{sec_res}

We compare the signal recovery performance from measurements that are optimized with the three proposed algorithms RCNCM-Elad, RCNCM-Xu and RCNCM-Duarte. The algorithms under test are: (i) random acquisition matrix with i.i.d. normal elements, (ii) the original Elad algorithm, (iii) original Xu algorithm, (iv) original Duarte algorithm, (v) proposed RCNCM-Elad algorithm, (vi) proposed RCNCM-Xu algorithm and (vii) proposed RCNCM-Duarte algorithm. As mentioned in Section \ref{subsec_Xu}, for the Xu and RCNCM-Xu algorithms we test a range of values of the step size $\alpha = 0.1, 0.2, ... 1.0$ and we consider an aggregate behaviour composed of the best results.

%

\begin{figure*}
  \centering
  \subfloat[Learned vs random dictionary]{
    \includegraphics[width=0.48\textwidth]{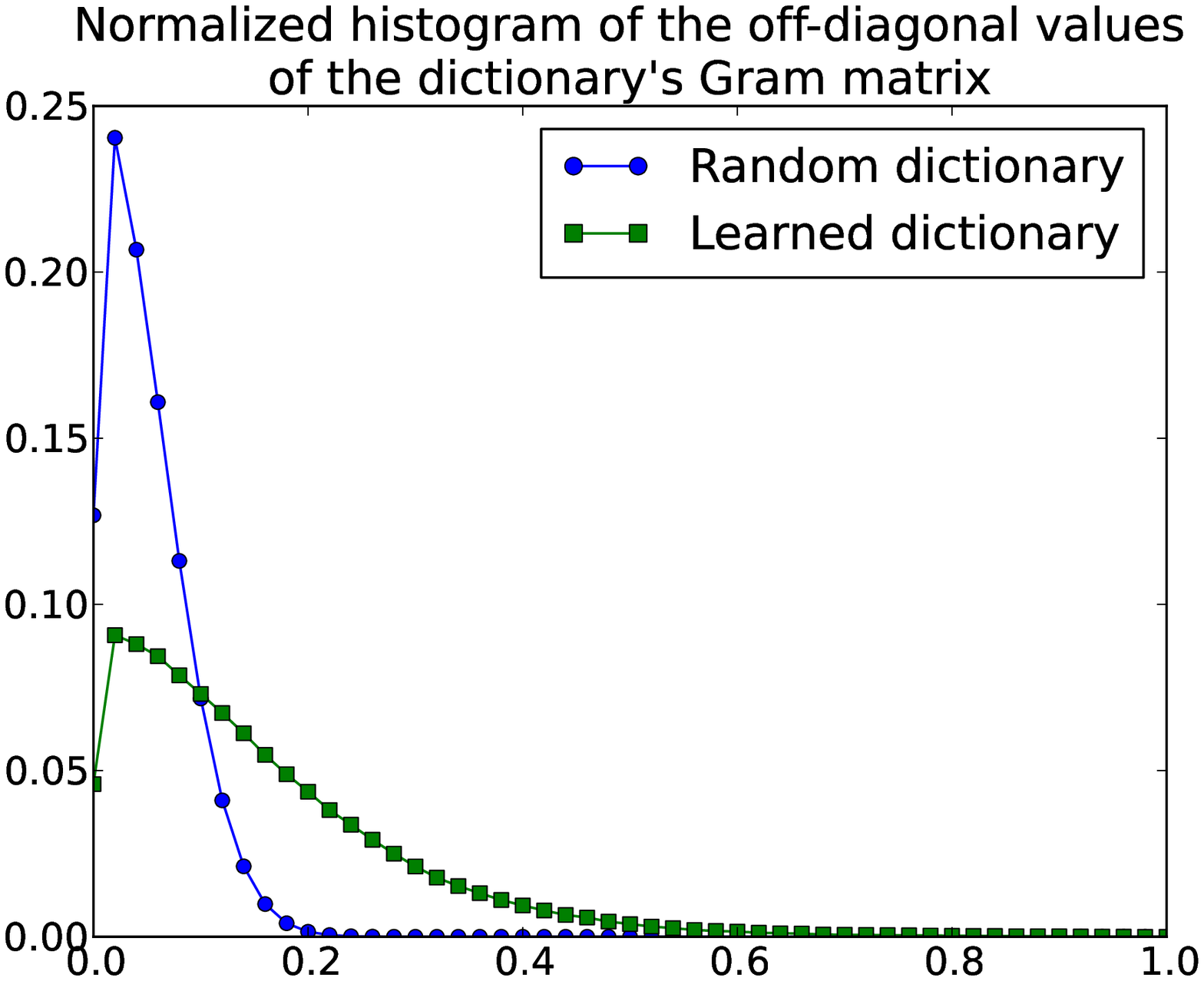}
    \label{fig_hists_test1}
  }
  \subfloat[Synthetic dictionaries]{
    \includegraphics[width=0.48\textwidth]{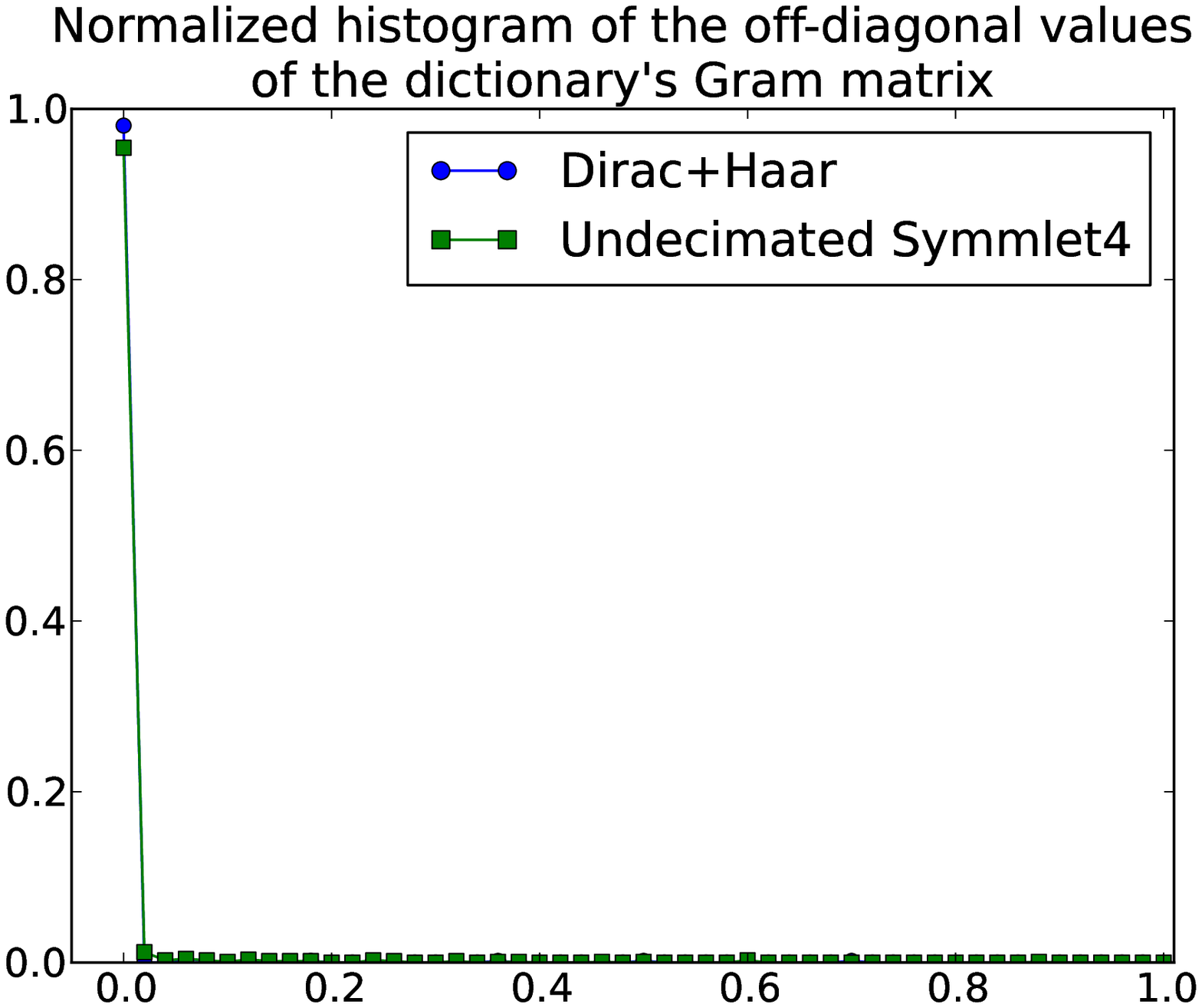}
    \label{fig_hists_test2}
  }
  \caption{Normalized histogram of the off-diagonal values of the dictionary's Gram matrix. The learned dictionary exhibits significant correlations, the synthetic dictionaries do not.}
  \label{fig_hists}
\end{figure*}

We use the K-SVD algorithm \cite{Aharon2006KSVD} to train a dictionary consisting of $N = 1024$ atoms for a set of randomly selected image patches.  The set is obtained by randomly selecting 150 patches of size $16 \times 16$ from each of 37 test images from the miscellaneous section of the public USC-SIPI image database \cite{USCSIPIdb}, resulting in a total of 5500 patches (there are 47 images in the database, but we removed some of them that were too uniform and affected the dictionary learning algorithm). The patches are reshaped columnwise as $256 \times 1$ vectors. The learned dictionary exhibits significant correlation between the atoms due to the similarities of the image patches, as shown in Fig.\ref{fig_hists_test1}, which depicts the histogram of the off-diagonal elements of the dictionary's Gram matrix compared to the histogram for a random dictionary of same size with i.i.d. normal elements. The existence of significant correlations implies that there is room for optimization of the projection vectors for better signal recovery.

We investigate the ratio of successful recovery of exact-sparse signals from $m = 150$ noisy measurements. The data is generated as random combinations of atoms from the dictionary. For reconstruction we use the following algorithms: (i) Orthogonal Matching Pursuit \cite{Pati1993} with stopping criterion being the residual error below a certain threshold $\epsilon$ given by the noise energy, denoted as \mbox{\emph{OMP-$\epsilon$}}, (ii) $\ell_1$ minimization, denoted as \emph{BP}, (iii) Robust Smoothed-$\ell_0$ \cite{SL02009,SL0robust} (SL0), (iv) Accelerated Iterative Hard Thresholding \cite{AIHT} (AIHT) and (v) Approximate Message Passing \cite{AMP} (AMP).

\begin{figure*}
  \centering
  \subfloat[OMP]{
  	\includegraphics[width=0.48\textwidth]{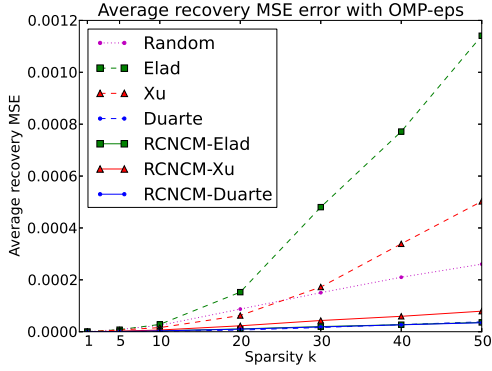}
  	\label{fig_test1_MSE_OMPeps}
  }
  \subfloat[BP]{
  	\includegraphics[width=0.48\textwidth]{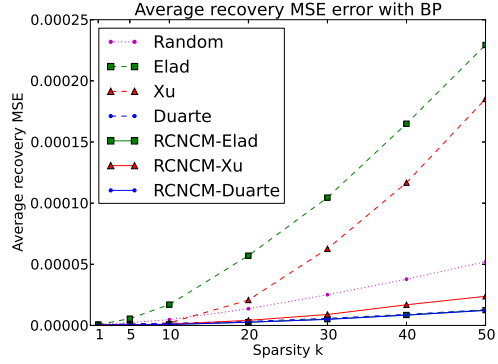}
  	\label{fig_test1_MSE_BP}
  }
  \\
  \subfloat[SL0]{
  	\includegraphics[width=0.48\textwidth]{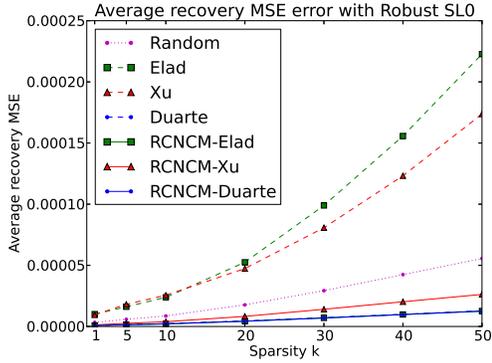}
  	\label{fig_test1_MSE_RobustSL0}
  }
  \subfloat[AIHT]{
  	\includegraphics[width=0.48\textwidth]{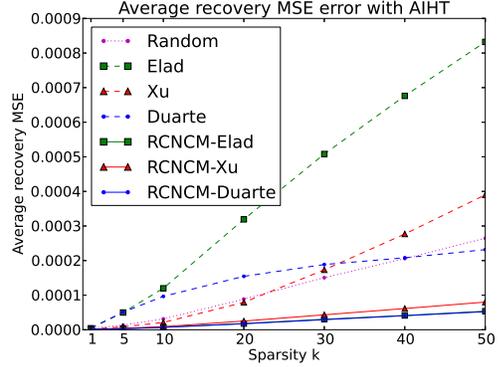}
  	\label{fig_test1_MSE_AIHT}
  }  
  \\
  \subfloat[AMP]{
  	\includegraphics[width=0.48\textwidth]{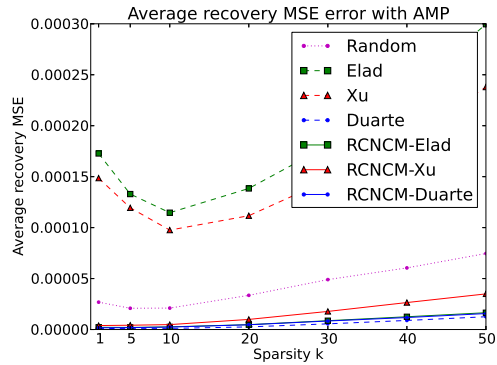}
  	\label{fig_test1_MSE_AMP}
  }
  \caption{Average MSE of reconstructed signals from $m=150$ optimized projections with $0.01$\% measurement noise energy (SNR=40dB). }
  \label{fig_learned_avgmse}  
\end{figure*}

Fig. \ref{fig_learned_avgmse} shows the average mean-squared-error (MSE) of the recovered signals for measurement signal-to-noise ratio (SNR) of 40dB ($0.01\%$ measurement noise). Note that the graph of  RCNCM-Elad is hidden under RCNCM-Duarte in all figures, as their performance is extremely close. Also the Duarte algorithm is superimposed with them in Fig.\ref{fig_test1_MSE_OMPeps}, Fig.\ref{fig_test1_MSE_BP} and Fig.\ref{fig_test1_MSE_RobustSL0} and thus not visible. 

The results show that the proposed algorithms consistently provide significantly lower recovery errors with all reconstruction algorithms, with RCNCM-Xu slightly behind RCNCM-Elad and RCNCM-Duarte. The original Duarte algorithm also matches their performance, except in the case of reconstruction with AIHT, in which it is distinctively worse. We have found no explanation for this particular behaviour of the Duarte algorithm with AIHT, but the behaviour was persistent in multiple simulations with different dictionaries created in the same manner in order to rule out the possibility of an error in the simulations.

The original Elad and Xu algorithms display poor performance in this test. Under different conditions (fewer number of measurements, smaller SNR), their relative performance is better, but still significantly behind Duarte and the proposed methods.

We remind the reader that the only difference between the original Elad and Xu algorithms and their proposed modifications RCNCM-Elad and RCNCM-Xu is that the minimization goal $\norm{G_e - I_N}$ is replaced with $\norm{G_e - G}$, a feature that is also shared by both the Duarte and RCNCM-Duarte algorithms that perform almost equally good. Thus, this simulation reveals the importance of taking into account the correlations between the atoms when they are expected to be significant, as in the case of the learned dictionary used (Fig.\ref{fig_hists}).

On the other hand, the difference between the Duarte and the RCNCM-Duarte algorithms consists in the normalization of the resulting atoms of the effective dictionary, which is not essential in this case. As the dictionary is learned and has no particular structure, the atoms of the Duarte effective dictionary, which is the top part of the right singular matrix of the dictionary (see section \ref{sec_Du_analysis}), have norms which do not vary greatly, and as such the performance of the two algorithms is similar. This explains why in Fig.\ref{fig_learned_avgmse} RCNCM-Duarte shows little improvement over the original Duarte algorithm (except in the case of AIHT). On the contrary, this additional normalization becomes essential with some highly structured dictionaries, as illustrated in the following section. 

\subsection*{The importance of the atom normalization constraint}
\label{subsec_synthdicts}

We also test the synthetic dictionaries listed in Section \ref{subsec_improvingDuarte} that pose a challenge to the Duarte algorithm because of the lack of normalization of the effective dictionary. The first is an orthogonal $256 \times 256$ dictionary with imprecise atom normalization varying between $1 \pm 10^{-6}$. The differences between atom norms are very small, but enough to make the dictionary's principal components (i.e. Duarte projection vectors) to be actual atoms of the dictionary, and as such they are orthogonal to all other atoms and fail to capture anything of them. The second dictionary is the concatenation of the Dirac and Haar orthonormal bases (size $256 \times 512)$, whereas the third is the non-decimated dictionary of a 2-level Symmlet4 wavelet decomposition, of size $256 \times 768$. The signal dimension is $n=256$.

\begin{figure*}
  \centering
  \subfloat[OMP]{
  	\includegraphics[width=0.48\textwidth]{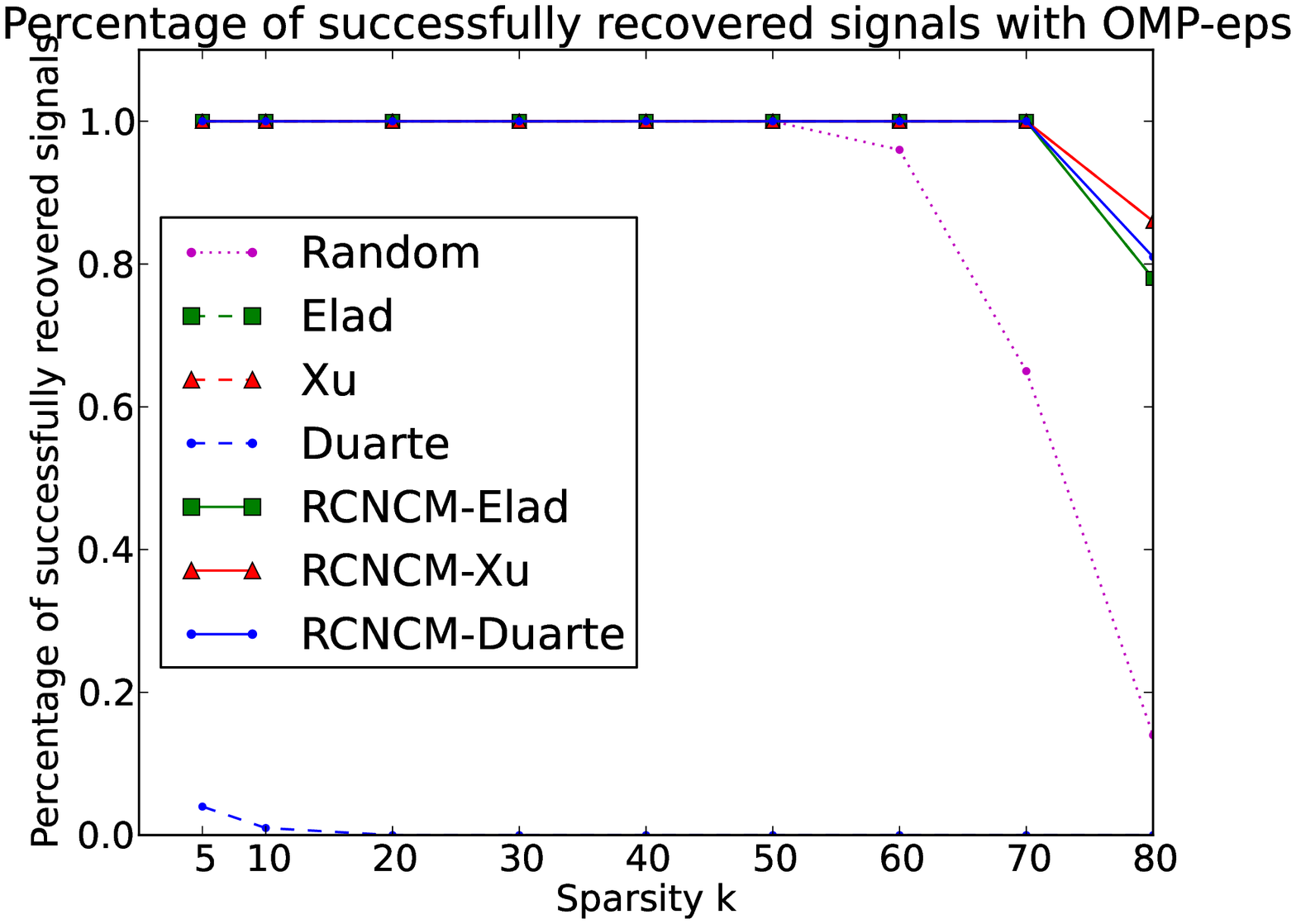}
  	\label{fig_test2_succprob_orthononnorm_OMPeps}
  }
  \subfloat[SL0]{
  	\includegraphics[width=0.45\textwidth]{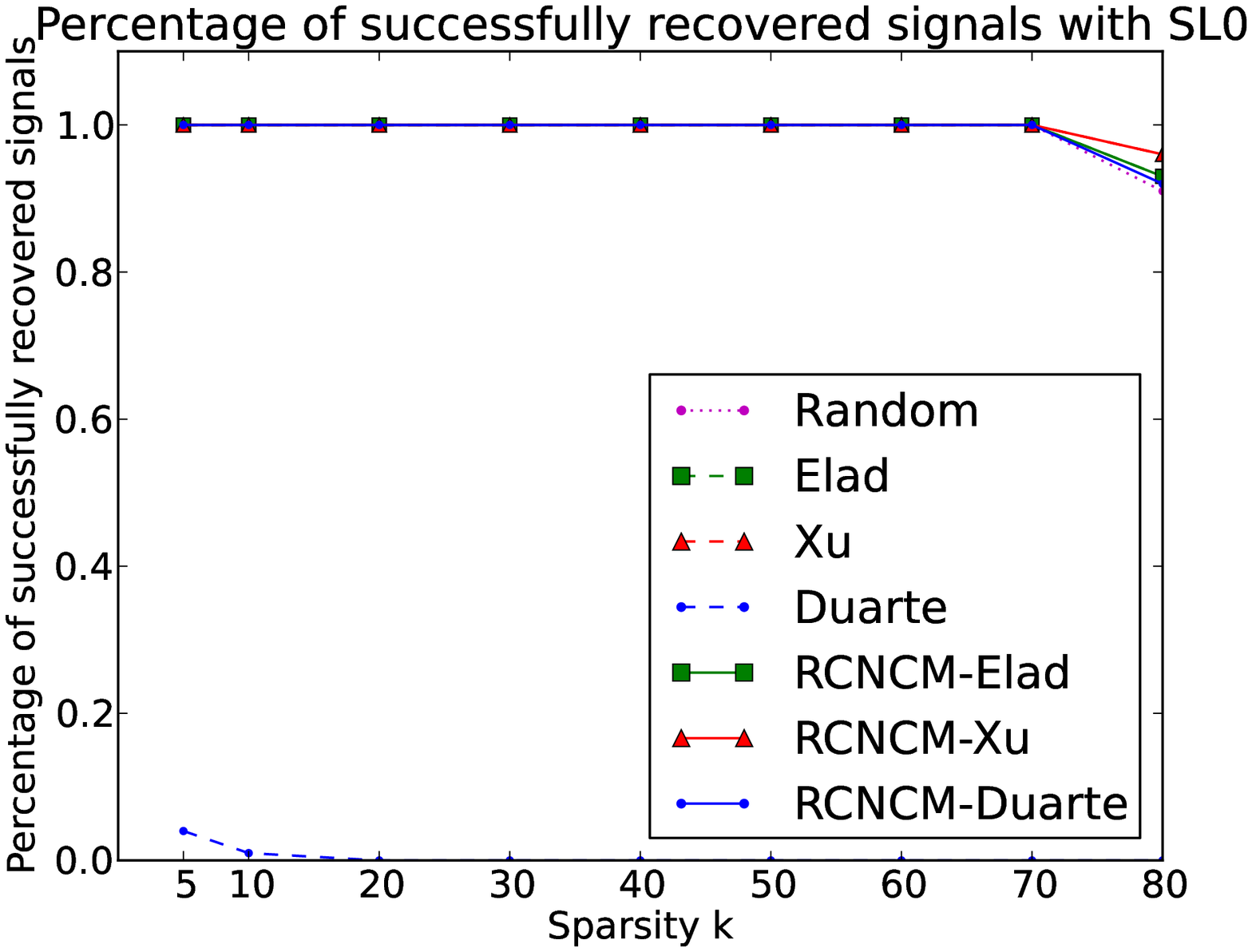}
  	\label{fig_test2_succprob_orthononnorm_SL0}
  }
  \\
  \subfloat[OMP]{
  	\includegraphics[width=0.48\textwidth]{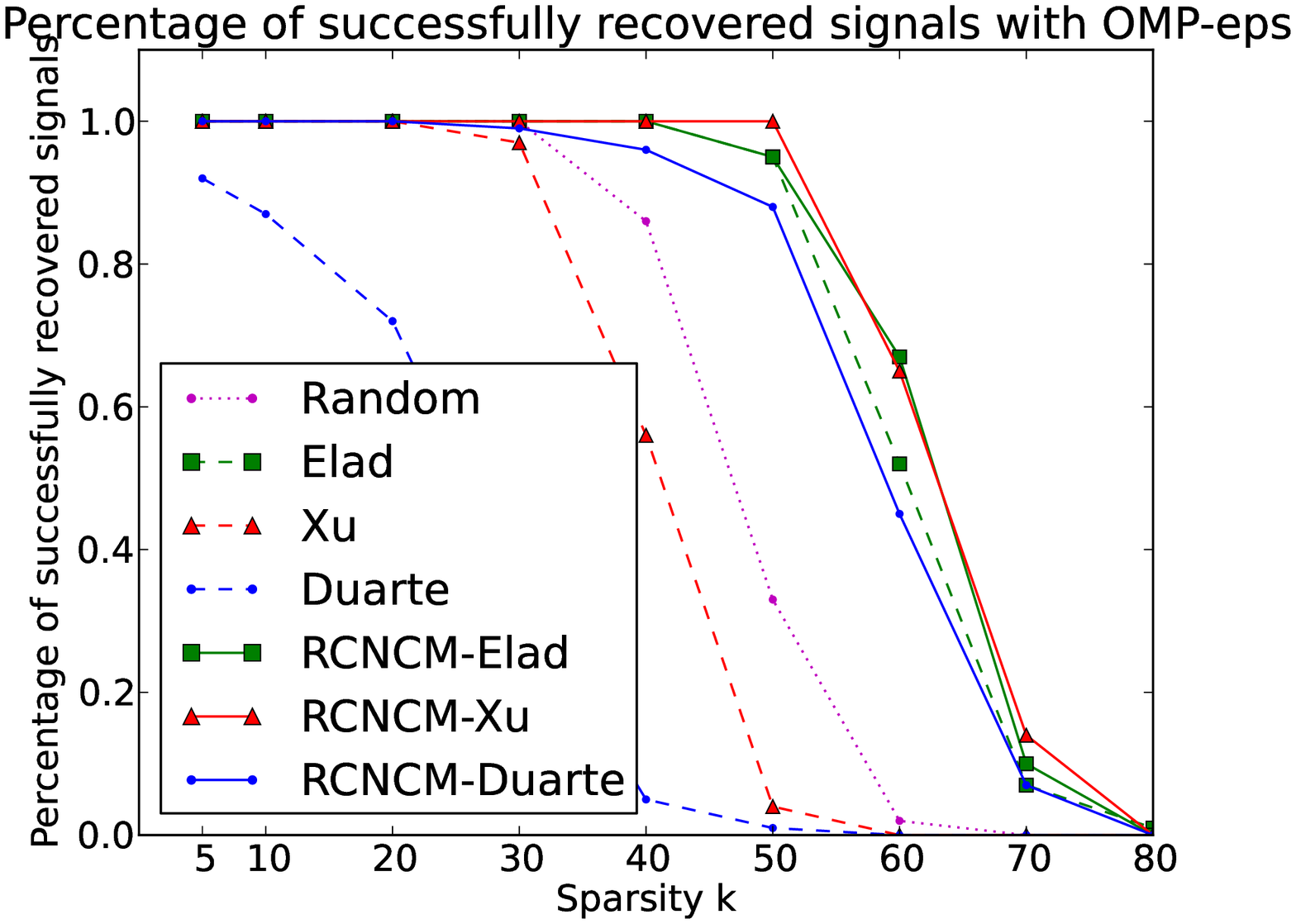}
  	\label{fig_test2_succprob_DiracHaar_OMPeps}
  }
  \subfloat[SL0]{
  	\includegraphics[width=0.45\textwidth]{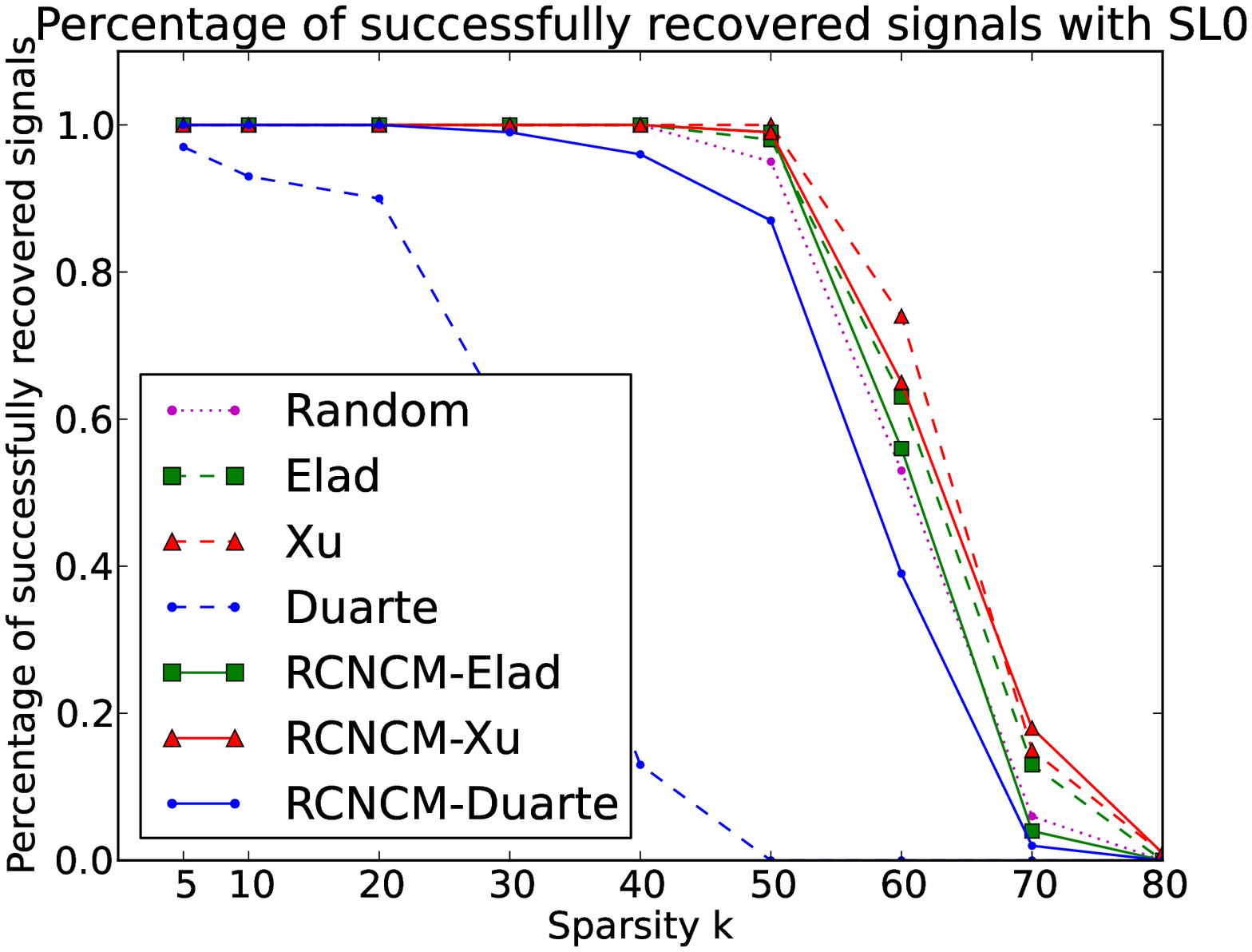}
  	\label{fig_test2_succprob_DiracHaar_SL0}
  }
  \\
  \subfloat[OMP]{
  	\includegraphics[width=0.48\textwidth]{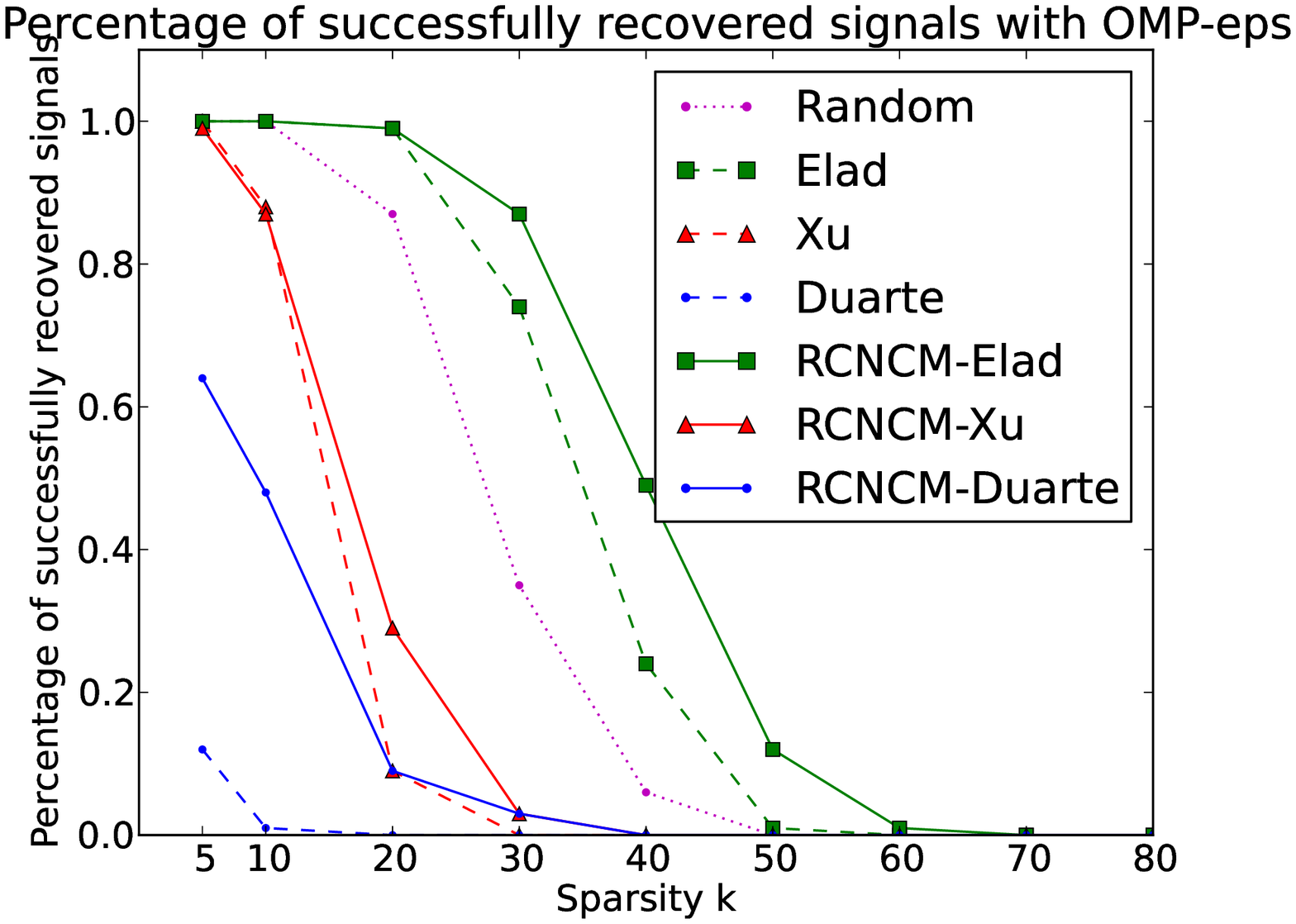}
  	\label{fig_test2_succprob_SWTsymm4_OMPeps}
  }
  \subfloat[SL0]{
  	\includegraphics[width=0.45\textwidth]{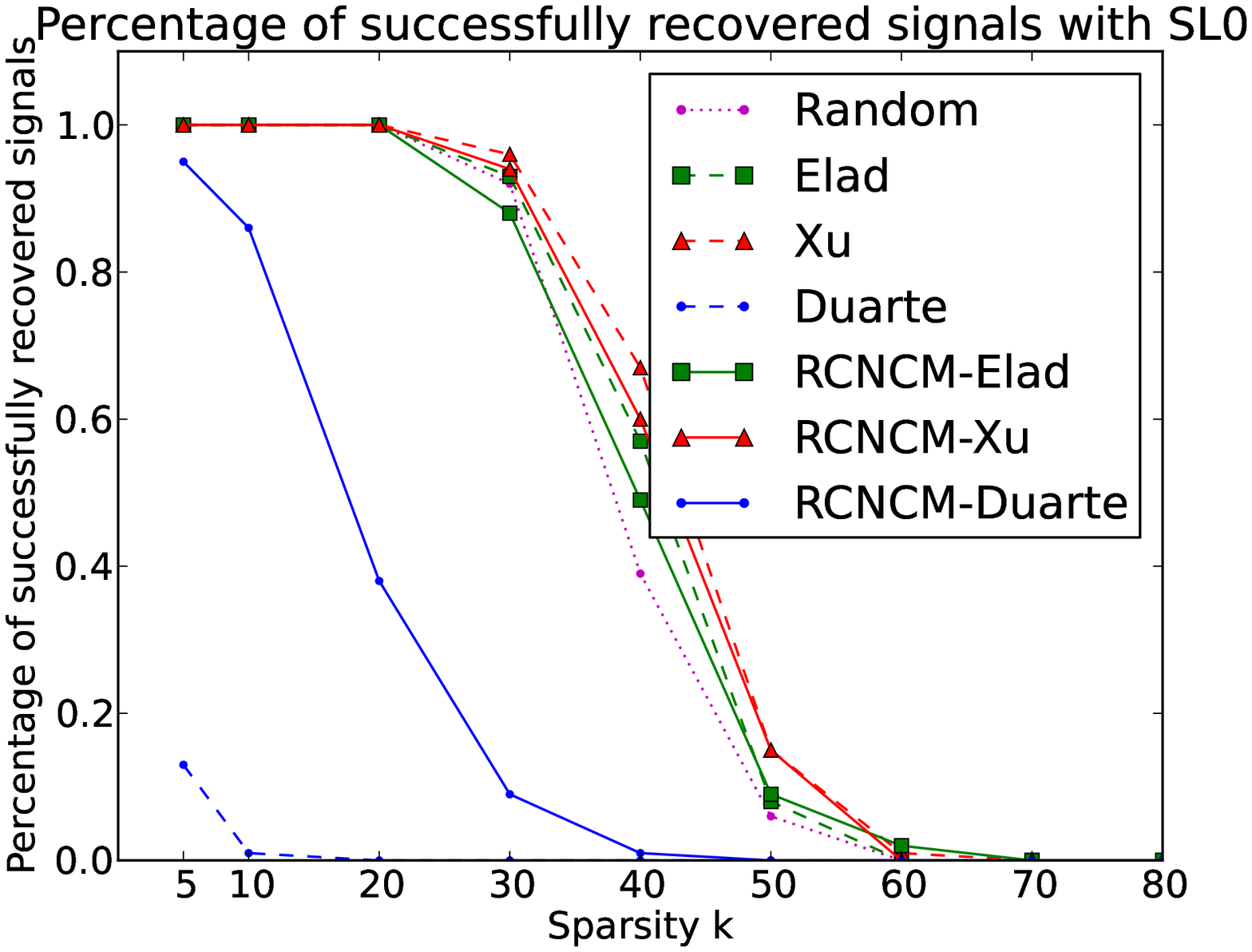}
  	\label{fig_test2_succprob_SWTsymm4_SL0}
  }  
  \caption{Percentage of successfully reconstructed exact-sparse data from $m=150$ noiseless optimized projections. }
  \label{fig_synth_successprob}  
\end{figure*}

Fig.\ref{fig_synth_successprob} shows the probability of successful reconstruction of exact-sparse data from $m=150$ noiseless measurements optimized with the algorithms under test. In the interest of brevity we only display the results obtained with OMP and SL0 recovery algorithms, which generally performed best. We consider a signal exactly recovered if the average MSE is smaller then $10^{-6}$. Please note that in Fig.\ref{fig_test2_succprob_orthononnorm_OMPeps} and \ref{fig_test2_succprob_orthononnorm_SL0} the Duarte graph is at the bottom of the figures.

As predicted, the Duarte algorithm is particularly sensitive to these highly structured dictionaries. The RCNCM-Duarte algorithm, which adds the unit-norm constraint, exhibits significant improvements over the Duarte algorithm, although in the case of the Symmlet4 dictionary is it still largely behind the other algorithms that all use the $\ell_\infty$ metric. We emphasize that in all our tests the Duarte algorithm performed very well with learned dictionaries, e.g. in Fig.\ref{fig_learned_avgmse}, and is only sensitive for some particularly unfavourable structured dictionaries like these. Thus in many cases it is a viable option, especially considering its simplicity (essentially, finding the principal components).

Note that in general the RCNCM-Elad and RCNCM-Xu algorithms do not provide significant improvements in this scenario, since the dictionaries do not have, on average, enough atom correlation: the first dictionary is actually orthogonal, whereas the other two have over 95\% of the Gram matrix off-diagonal elements smaller in absolute value than $0.01$, as depicted in Fig.\ref{fig_hists_test2}. This is due to their construction from orthogonal bases (concatenating two orthonormal and very sparse bases, or the undecimated version of an orthonormal wavelet transform). The behavior is contrary to the previous scenario, as learned dictionaries generally exhibit atom correlations but no particular structure in the right singular matrix, and therefore no great differences in the norms of the Duarte effective dictionary. As such, learned dictionaries benefit mostly from the RCNCM-Elad / Xu algorithms but not from RCNCM-Duarte, and the synthetic dictionaries benefit oppositely. In order to have improvements for all three proposed algorithms at once, the dictionary should simultaneously have significant atom correlations and a sparse or concentrated structure of the right singular matrix. However we encountered no such dictionary in our simulations, as the two features seem to be of opposite nature.



Considering the recovery results with both the learned and the synthetic dictionaries, we conclude that the proposed algorithms provide better measurements than their originals in relevant scenarios, increasing the accuracy and robustness of the recovery process. In particular, the proposed RCNCM-Elad algorithm consistently provided very good results with all recovery algorithms and dictionaries we tested on.

\section{Conclusions}
\label{sec_concl}

This paper focuses on optimizing the acquisition matrix for compressive sensing of signals that are sparse in overcomplete dictionaries. We propose improvements for three existing optimization algorithms, based on analyzing them in particular cases where they perform sub-optimally. We argue that the Elad and Xu optimization algorithms can be improved by making the effective dictionary mimic the correlations of the original dictionary, instead of just reducing its mutual coherence. To the Duarte algorithm we propose adding an additional unit-norm constraint to the optimization problem, in order to avoid having atoms with small norms in the effective dictionary.

Furthermore, all the three modified algorithms can be viewed as special instances of a single unified formulation, in the form of a rank-constrained nearest correlation matrix problem: find the matrix that is of minimal distance from the Gram matrix of the dictionary, subject to rank, unit-norm and semipositivity constraints. When the distance metric is the $\ell_2$ norm, the problem becomes similar to the Duarte optimization problem with the proposed additional unit-norm constraint. Several algorithms have already been developed for this optimization problem. When the distance metric is $\ell_\infty$, one can use the modified Elad and Xu algorithms, iteratively minimizing the largest entries of the difference of Gram matrices.

Simulation results show increased signal recovery accuracy, as well as better robustness with particular structured dictionaries that the Duarte algorithm is sensitive to. We conclude that formulating the optimization problem as a rank-constrained nearest correlation matrix problem is more accurate and robust than the existing approaches for optimizing the acquisition matrix.

\section{Acknowledgments}
\label{sec_ack}

This paper was partly realized with the support of EURODOC ``Doctoral Scholarships for research performance at European level'' project, financed by the European Social Found and Romanian Government.

The author would like to thank professor Liviu Gora\c{s} for his guidance and advice.

\section*{References}

\bibliographystyle{elsarticle-num} 

\end{document}